\documentclass[aps,twocolumn,amsmath,amssymb,prl,superscriptaddress,showpacs]{revtex4}
\usepackage{graphicx}
\usepackage{dcolumn}
\usepackage{bm}
\usepackage{color}

\begin{document}

\title{Proximity-free enhancement of anomalous Nernst effects in metallic multilayers}
\author{Ken-ichi Uchida}
\email{kuchida@imr.tohoku.ac.jp}
\affiliation{Institute for Materials Research, Tohoku University, Sendai 980-8577, Japan}
\affiliation{PRESTO, Japan Science and Technology Agency, Saitama 332-0012, Japan}
\author{Takashi Kikkawa}
\affiliation{Institute for Materials Research, Tohoku University, Sendai 980-8577, Japan}
%
\author{Takeshi Seki}
\affiliation{Institute for Materials Research, Tohoku University, Sendai 980-8577, Japan}
\author{Takafumi Oyake}
\affiliation{Department of Mechanical Engineering, The University of Tokyo, Tokyo 113-8656, Japan}
\author{Junichiro Shiomi}
\affiliation{Department of Mechanical Engineering, The University of Tokyo, Tokyo 113-8656, Japan}
\author{Koki Takanashi}
\affiliation{Institute for Materials Research, Tohoku University, Sendai 980-8577, Japan}
\author{Eiji Saitoh}
\affiliation{Institute for Materials Research, Tohoku University, Sendai 980-8577, Japan}
\affiliation{Spin Quantum Rectification Project, ERATO, Japan Science and Technology Agency, Sendai 980-8577, Japan}
\affiliation{WPI Advanced Institute for Materials Research, Tohoku University, Sendai 980-8577, Japan}
\affiliation{Advanced Science Research Center, Japan Atomic Energy Agency, Tokai 319-1195, Japan}
\date{\today}
\begin{abstract}
The anomalous Nernst effect (ANE) has been investigated in alternately-stacked multilayer films comprising paramagnetic and ferromagnetic metals. We found that the ANE is enhanced with increasing the number of the paramagnet/ferromagnet interfaces with keeping the total thickness of the films constant, and that the enhancement appears even in the absence of magnetic proximity effects; similar behavior was observed not only in Pt/Fe multilayers but also in Au/Fe and Cu/Fe multilayers free from proximity ferromagnetism. This universal enhancement of the ANE in the metallic multilayers suggests the presence of unconventional interface-induced thermoelectric conversion in the Fe films attached to the paramagnets. 
\end{abstract}
\maketitle
%
%
The anomalous Nernst effect (ANE) is one of the transverse thermoelectric effects in ferromagnetic materials \cite{ANE_Berger,ANE_Miyasato,ANE_Mizuguchi,ANE_Weischenberg,ANE_Ramos,ANE_Sakuraba,spintronics2}. The electric field induced by the ANE ${\bf E}_{\rm ANE}$ is generated via spin-orbit interaction in the direction of the cross product of the spontaneous magnetization ${\bf M}$ and applied temperature gradient $\nabla T$:  
\begin{equation}\label{equ:ANE1}
{\bf E}_{\rm ANE} = S_{\rm ANE} \nabla T \times \left( { \frac{\bf M}{|{\bf M}|} } \right), 
\end{equation}
where $S_{\rm ANE}$ is the anomalous Nernst coefficient. Although the ANE is a well-known phenomenon having a long research history, it is drawing renewed attention in the field of spintronics \cite{spintronics1,spintronics2}. From the viewpoint of fundamental physics, key targets in the ANE research include microscopic understanding of the mechanism of this phenomenon \cite{ANE_Miyasato,ANE_Weischenberg,ANE_Ramos} and separation of the ANE from the spin Seebeck effects (SSEs) \cite{TSSE_Uchida2008,TSSE_Uchida2010,TSSE_Jaworski2010,LSSE_Uchida2010,LSSE_Du2013PRL,LSSE_Kikkawa2013PRL,LSSE_Schreier2013,LSSE_Kikkawa2013PRB,LSSE_Rezende2014,LSSE-UchidaJPCM,LSSE_Uchida2013PRB,LSSE_Ramos2013}. From the viewpoint of applications, development of novel thermoelectric generation technology based on the ANE is already in progress \cite{ANE_Sakuraba}. \par
Recently, the ANE has been investigated also in paramagnetic metals connected to ferromagnetic materials for revealing the effect of magnetic proximity on thermal spin-transport phenomena \cite{Huang_2012PRL,Lu_2013PRL}. In a paramagnet/ferromagnet junction system, when the paramagnet is near the Stoner ferromagnetic instability (e.g., Pt and Pd) \cite{Ibach,DOS}, ferromagnetism may be induced in the paramagnet in the vicinity of the paramagnet/ferromagnet interface due to static magnetic proximity effects. If the proximity ferromagnetism is combined with spin-orbit interaction, the ANE may appear even in the paramagnetic materials. In 2012, Huang {\it et al.} \cite{Huang_2012PRL} pointed out that the proximity-induced ANE (PANE) might contaminate the longitudinal SSE (LSSE) \cite{LSSE_Uchida2010,LSSE_Du2013PRL,LSSE_Kikkawa2013PRL,LSSE_Schreier2013,LSSE_Kikkawa2013PRB,LSSE_Rezende2014,LSSE-UchidaJPCM,LSSE_Uchida2013PRB,LSSE_Ramos2013} in Pt/Y$_3$Fe$_5$O$_{12}$ (YIG) junction systems, which are commonly used for investigating spin-current phenomena. Followed by this problem presentation, we experimentally demonstrated that transverse thermoelectric voltage in Pt/YIG systems is due purely to the LSSE and established a method for the clear separation of the PANE from the LSSE \cite{LSSE_Kikkawa2013PRL,LSSE_Kikkawa2013PRB,LSSE-UchidaJPCM}. In 2014, Guo {\it et al.} \cite{ANE_Guo} theoretically investigated the PANE in ferromagnetic Pt and Pd within Berry-phase formalism based on relativistic band-structure calculations; the magnitude of the PANE coefficient for Pt on YIG was predicted to be small: $S_{\rm ANE} \sim 0.06~\mu \textrm{V/K}$. The PANE contribution to the output voltage in real Pt/YIG systems is further reduced because of short-circuit effects; since the proximity ferromagnetism in Pt exists only in several atomic layers adjacent to the Pt/YIG interface and the remaining region is a paramagnetic metal with high electrical conductivity, the PANE in the thin ferromagnetic region is electrically shunted \cite{LSSE_Kikkawa2013PRB,LSSE-UchidaJPCM}. Due to this situation, there is no clear evidence for the existence of the PANE; only the upper limit of the PANE contribution in the Pt/YIG systems is provided \cite{LSSE_Kikkawa2013PRB}. \par
\begin{figure}[ht]
\begin{center}
\includegraphics{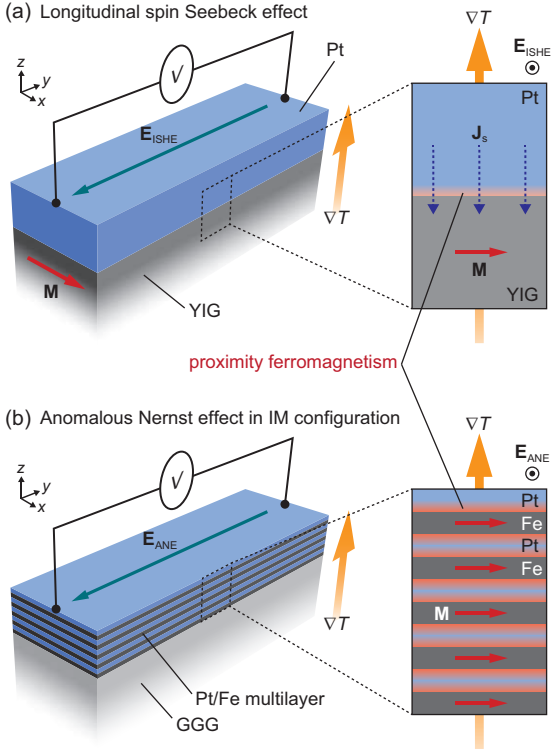}
\caption{Schematic illustrations of the LSSE in the Pt/YIG junction (a) and the ANE in the IM configuration in the Pt/Fe multilayer sample formed on the GGG substrate (b). $\nabla T$, ${\bf M}$, ${\bf J}_{\rm s}$, ${\bf E}_{\rm ISHE}$, and ${\bf E}_{\rm ANE}$ denote the temperature gradient, magnetization vector, spatial direction of the thermally generated spin current, electric field induced by the ISHE, and electric field induced by the ANE, respectively. Red areas in the Pt layers schematically illustrate proximity-induced ferromagnetic regions. }\label{fig:1}
\end{center}
\end{figure}
Then, is it possible to enhance the PANE intentionally? A straightforward way to answer this question is to increase the density of proximity ferromagnetism, since the ANE is proportional to the magnetization in general \cite{ANE_Berger}. This is realized by attaching paramagnets to ferromagnets with large saturation magnetization and by increasing the number of the paramagnet/ferromagnet interfaces per unit volume. Following this strategy, we focus on Pt/Fe multilayers, where the saturation magnetization of Fe ($\sim$ 22.1 kG) \cite{ANE_Weischenberg} is 12 times greater than that of YIG ($\sim$ 1.8 kG) \cite{Lu_2013PRL,Geprags_2012APL} at room temperature. By stacking thin Pt/Fe films, the density of the proximity-induced Pt ferromagnetism in the Pt/Fe multilayers can be much greater than that in the conventional Pt/YIG systems (see Fig. \ref{fig:1}). In fact, proximity-induced magnetic moments in Pt connected to Fe were confirmed to be much greater than those in Pt connected to YIG by the measurements of X-ray magnetic circular dichroism \cite{Geprags_2012APL}. In this study, to investigate the possible enhancement of the PANE, we measured the transverse thermoelectric voltage in the Pt/Fe multilayers in different magnetization and temperature-gradient configurations with changing the Pt/Fe-interface density. \par
\begin{figure*}[ht]
\begin{center}
\includegraphics{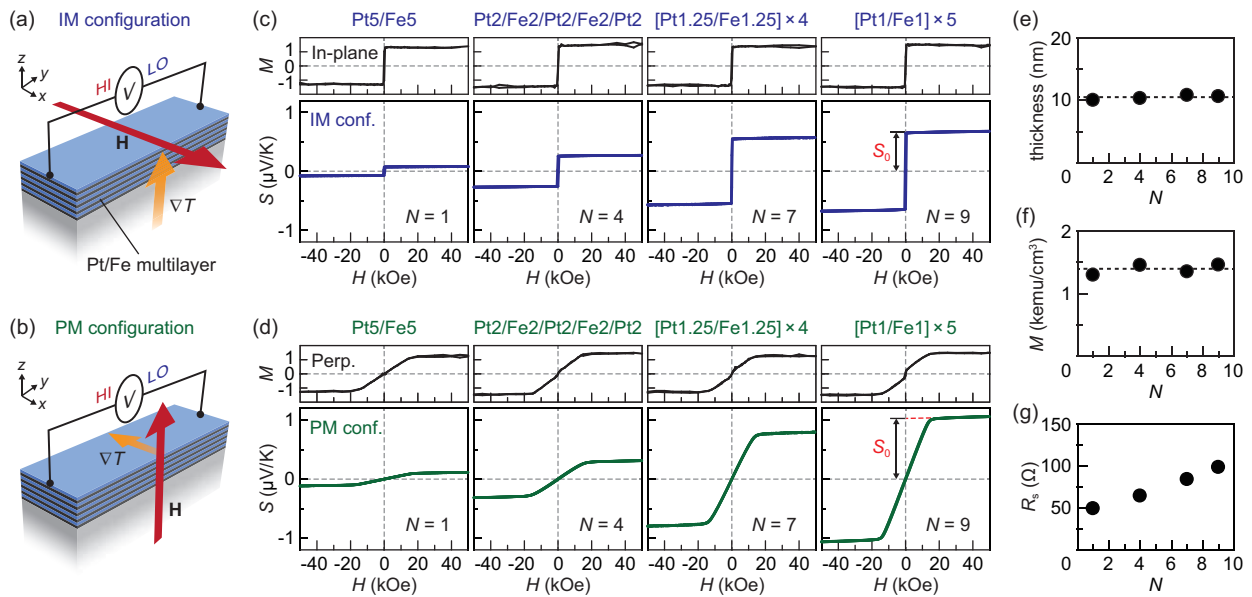}
\caption{(a),(b) Schematic illustrations of the Pt/Fe multilayer sample in the IM (a) and PM (b) configurations. ${\bf H}$ denotes the magnetic field vector with the magnitude of $H$. During the measurements of the ANE, a uniform external magnetic field was applied by using a superconducting solenoid magnet. (c) The in-plane magnetization curve and the $H$ dependence of the transverse thermopower $S$ for the Pt/Fe multilayer samples for various values of $N$ in the IM configuration. $M$ and $N$ denote the total magnetization (in units of $\textrm{kemu/cm}^3$) of the Pt/Fe multilayer samples, measured with a vibrating sample magnetometer, and the number of the Pt/Fe interfaces, respectively. The $M$ data were measured by using Pt/Fe multilayer films formed on thermally oxidized silicon substrates since the large paramagnetism of GGG is an obstacle to detect $M$ of the thin films. In the $M$-$H$ curves, the contributions from the silicon substrates are subtracted. $S_0$ is the anomalous component of $S$, obtained by extrapolating the $S$-$H$ curve in the high $H$ field range to zero field. (d) The perpendicular magnetization curve and the $H$ dependence of $S$ for the Pt/Fe multilayer samples for various values of $N$ in the PM configuration. (e) $N$ dependence of the total thickness of the Pt/Fe multilayer samples, measured with a surface profiler. (f) $N$ dependence of $M$. (g) $N$ dependence of the sheet resistance $R_{\rm s}$ of the Pt/Fe multilayer samples in the $y$ direction, measured by a four-probe method. }\label{fig:2}
\end{center}
\end{figure*}

Figure \ref{fig:1}(b) shows the schematic illustration of the Pt/Fe multilayer sample used in the present study. We prepared four alternately-stacked Pt/Fe multilayer samples with the different layer number: Pt5/Fe5, Pt2/Fe2/Pt2/Fe2/Pt2, [Pt1.25/Fe1.25]$\times$4, and [Pt1/Fe1]$\times$5 samples, where the number of the Pt/Fe interfaces is $N = 1$, 4, 7, and 9, respectively, and the numeric characters after Pt and Fe represent the thicknesses in units of nm; for example, the Pt5/Fe5 sample refers to a bilayer film comprising a 5-nm-thick Pt film and a 5-nm-thick Fe film. Importantly, the total thickness of all the multilayer films is fixed at 10 nm [Fig. \ref{fig:2}(e)]; the layer density of the Pt/Fe interfaces monotonically increases with increasing $N$, a situation different from other experiments \cite{ANE-multilayer1,SSE-multilayer}. The Pt/Fe multilayer films were formed on the whole surface of single-crystalline Gd$_3$Ga$_5$O$_{12}$ (GGG) (111) substrates by means of ultrahigh vacuum magnetron sputtering at ambient temperature. The substrates do not affect the ANE in the Pt/Fe multilayers because GGG is a paramagnetic insulator \cite{comment-PSSE,paramagnetic-SSE}. Since the top layer of all the Pt/Fe multilayers is Pt, we can exclude possible artifacts caused by the oxidation of the Fe layers. The lengths of the GGG substrates along the $x$, $y$, and $z$ directions are $L_x = 2~\textrm{mm}$, $L_y = 6~\textrm{mm}$, and $L_z = 1~\textrm{mm}$, respectively. We checked that the total magnetization of the Pt/Fe multilayer films is almost independent of $N$ [Fig. \ref{fig:2}(f)], indicating that the saturation magnetization of each Fe layer is not changed in these multilayer films. Owing to the large saturation magnetization of Fe and the high Pt/Fe-interface density, the PANE in the Pt/Fe multilayer sample with $N=9$ is expected to be two orders of magnitude greater than that in the conventional Pt (10 nm)/YIG systems. \par
The ANE in the Pt/Fe multilayer samples was measured in two different configurations. One is an in-plane magnetized (IM) configuration, in which an external magnetic field ${\bf H}$ with the magnitude $H$ is applied parallel to the Pt/Fe interface and a temperature gradient $\nabla T$ is applied perpendicular to the interface [see Fig. \ref{fig:2}(a)]. The IM configuration is widely used for measuring the LSSE in paramagnet/ferromagnet junction systems, where the symmetry of the LSSE is determined by the inverse spin Hall effect (ISHE) \cite{ISHE1,ISHE2,ISHE3} in the paramagnet: 
\begin{equation}\label{equ:SSE1}
{\bf E}_{\rm ISHE} = D_{\rm ISHE} {\bf J}_{\rm s} \times {\bm \sigma}
\end{equation}
with $D_{\rm ISHE}$, ${\bf E}_{\rm ISHE}$, ${\bf J}_{\rm s}$, and ${\bm \sigma}$ respectively being the ISHE coefficient, electric field induced by the ISHE, spatial direction of the spin current flowing perpendicular to the paramagnet/ferromagnet interface, and spin-polarization vector in the paramagnet parallel to ${\bf M}$ of the adjacent ferromagnet. Since ${\bf J}_{\rm s}$ is parallel to $\nabla T$, both the LSSE and ANE can exhibit the transverse thermoelectric voltage in the IM configuration [see Eqs. (\ref{equ:ANE1}) and (\ref{equ:SSE1})]. The other is a perpendicularly magnetized (PM) configuration, in which ${\bf H}$ is perpendicular to the Pt/Fe interface and $\nabla T$ is parallel to the interface [see Fig. \ref{fig:2}(b)]. In the PM configuration, the ANE voltage can appear, while the LSSE voltage disappears due to the symmetry of the ISHE [see Eq. (\ref{equ:SSE1}) and note that ${\bf J}_{\rm s}~||~{\bm \sigma}$ in the PM configuration] \cite{comment-PM}. Therefore, the comparison of the transverse thermoelectric voltage between the IM and PM configurations enables the separation of the ANE from the LSSE \cite{LSSE_Kikkawa2013PRL,LSSE_Kikkawa2013PRB,LSSE-UchidaJPCM}. To generate $\nabla T$ in the IM and PM configurations, the samples were sandwiched between two highly thermally conductive AlN plates, of which the temperatures are stabilized at $300~\textrm{K}$ and $300~\textrm{K} + \Delta T$ with the temperature difference $\Delta T$. The details of the experimental procedure are described in Ref. \onlinecite{LSSE_Kikkawa2013PRB}. We measured an electric voltage difference $V$ between the end of the Pt/Fe multilayer samples along the $y$ direction. Hereafter, we mainly plot the transverse thermopower $S \equiv (V/\Delta T)(L_z / L_{y})$ [$S \equiv (V/\Delta T)(L_x / L_{y})$] in the IM (PM) configuration. \par
\begin{figure*}[ht]
\begin{center}
\includegraphics{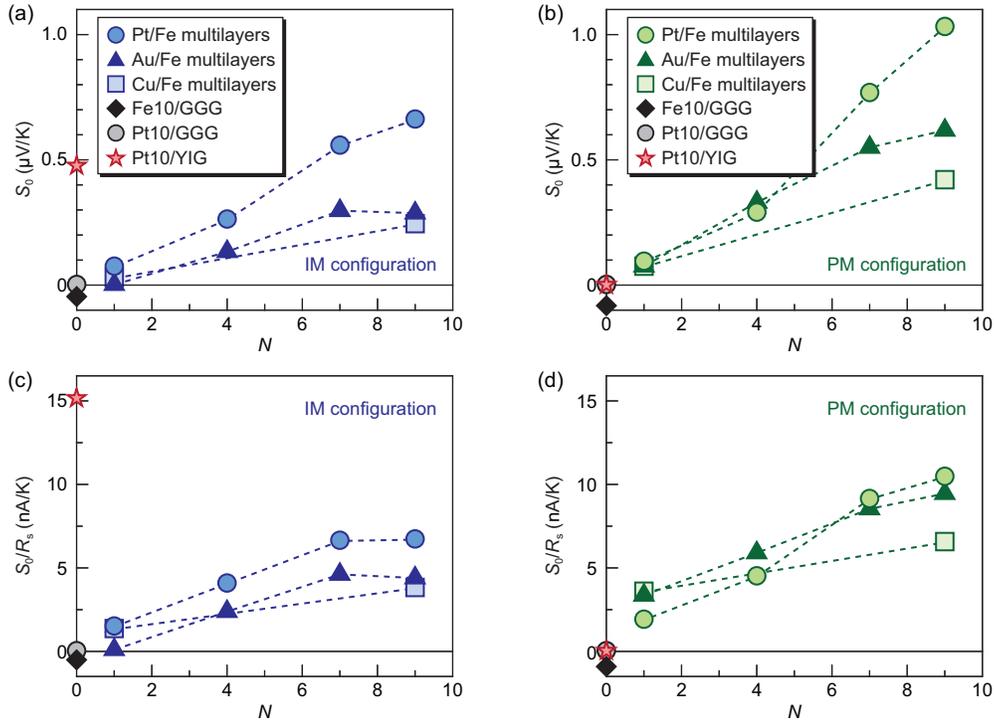}
\caption{(a),(b) $N$ dependence of $S_0$ in the Pt/Fe, Au/Fe, and Cu/Fe multilayer samples in the IM (a) and PM (b) configurations. The total thickness of these multilayers is 10 nm. The $S_0$ signals in a 10-nm-thick Fe film on the GGG substrate and in 10-nm-thick Pt films on the GGG and YIG substrates are plotted at $N=0$. The Cu/Fe multilayer samples are covered with 1-nm-thick Pt films to avoid the oxidation of the top Cu layers. (c),(d) $N$ dependence of $S_0/R_{\rm s}$ in the Pt/Fe multilayer, Au/Fe multilayer, Cu/Fe multilayer, Fe/GGG, Pt/GGG, and Pt/YIG samples in the IM (c) and PM (d) configurations. }\label{fig:3}
\end{center}
\end{figure*}

In Fig. \ref{fig:2}(c), we compare the $H$ dependence of $S$ with the in-plane magnetization curve for the Pt/Fe multilayer films. All the Pt/Fe multilayers exhibit the clear $S$ signals of which the sign is reversed in response to their magnetization reversal. We found that the magnitude of the $S$ signals monotonically increases with increasing the number of the Pt/Fe interfaces $N$. As shown in Fig. \ref{fig:2}(d), similar behavior was observed also in the PM configuration. This situation is completely different from the LSSE in the conventional Pt/YIG systems, where the $S$ signal in the PM configuration is negligibly small while that in the IM configuration is comparable in magnitude to the ANE in the Pt/Fe multilayers with the large $N$ values (see Refs. \onlinecite{LSSE_Kikkawa2013PRB,LSSE-UchidaJPCM} and red-star data points in Fig. \ref{fig:3}). Therefore, we can conclude that the ANE is enhanced with increasing $N$ in the Pt/Fe multilayer samples. We also confirmed that the $H$ dependence of $S$ is not affected by the magnetoresistance of the Pt/Fe multilayer samples, where the magnetoresistance ratio was observed to be much smaller than $<1~\%$ even for $N=9$. \par
The blue (green) circle data points in Fig. \ref{fig:3}(a) [\ref{fig:3}(b)] show the anomalous component of the $S$ signal, $S_{\rm 0}$, in the Pt/Fe multilayer samples as a function of $N$ in the IM (PM) configuration, where the $S_{\rm 0}$ values are extracted by extrapolating the $S$ data in the high field range ($50~\textrm{kOe}<H<90~\textrm{kOe}$) to zero field \cite{comment-ANEcoefficient}. The magnitude of $S_{\rm 0}$ increases roughly in proportion to $N$ in the Pt/Fe multilayer samples in both the configurations. We found that the $S_{\rm 0}$ signal normalized by the sheet resistance $R_{\rm s}$ shown in Fig. \ref{fig:2}(g), which represents the charge current generated by the ANE per unit temperature difference, also increases with increasing $N$ in the Pt/Fe multilayer samples [see Figs. \ref{fig:3}(c) and \ref{fig:3}(d)]. The strong correlation between the ANE and the Pt/Fe-interface density is seemingly consistent with the appearance of the PANE in the Pt layers, of which the magnitude should be enhanced with increasing $N$ as discussed above. \par
To clarify the origin of the ANE enhancement in the Pt/Fe multilayers, we performed the same measurements using Au/Fe and Cu/Fe multilayer samples, in which the Pt layers are replaced with Au and Cu films, respectively. Since Au and Cu are typical metals far from the Stoner instability \cite{Ibach,DOS} and proximity-induced magnetic moments in Au and Cu are much smaller than those in Pt \cite{Geprags_2012APL,XMCD_FeCu,XMCD_FePt,Calc_FePt_FeAu,Calc_FeCu,XMCD_CoAu}, the ANE measurements using these multilayers allow us to judge whether or not the observed enhancement of the ANE in the Pt/Fe multilayers originates from the proximity ferromagnetism in Pt. In Figs. \ref{fig:3}(a) and \ref{fig:3}(b), we also show the $N$ dependence of $S_0$ in the Au/Fe multilayer (closed triangles) and Cu/Fe multilayer (open squares) samples in the IM and PM configurations, respectively. We found that, even in the Au/Fe and Cu/Fe multilayer samples, the clear $S_0$ signals appear and their magnitude increases with increasing $N$. Although the magnitude of the $S_0$ signals in the Pt/Fe multilayers is greater than that in the Au/Fe and Cu/Fe multilayers, the difference in $S_0$ between these samples is attributed mainly to that in the sheet resistance; the magnitude of $S_0/R_{\rm s}$ in all the samples is comparable at each $N$ value [see Figs. \ref{fig:3}(c) and \ref{fig:3}(d)]. Here we note that $S_0/R_{\rm s}$ is more essential to compare the ANEs in these multilayer systems since this factor includes the correction coming from the short-circuit effects approximately. Therefore, we conclude that there is no convincing evidence for the existence of the PANE even in the Pt/Fe multilayers and that the main contribution to the ANE enhancement is not the magnetic proximity effect. If we assume that the difference in the $S_0/R_{\rm s}$ values between the Pt/Fe and Au/Fe multilayers comes from the proximity ferromagnetism, the upper limit of the PANE contribution in the Pt/Fe multilayer sample with $N=9$ is estimated to be $S_0/R_{\rm s} = 2.3~\textrm{nA/K}$ ($1.0~\textrm{nA/K}$) in the IM (PM) configuration. However, the upper limit for the IM configuration cannot be interpreted quantitatively because of the possible contributions of the ISHE induced by thermal spin currents and of the correction factor for the thermal conductivity \cite{LSSE_Kikkawa2013PRB,LSSE-UchidaJPCM}. The upper limit for the PM configuration estimated here is not conflict with the PANE coefficient in Pt predicted by Guo {\it et al.} \cite{ANE_Guo}. \par
The above experiments demonstrate that the ANE is enhanced in the alternately-stacked paramagnet/ferromagnet multilayer films even in the absence of the proximity ferromagnetism in Pt. However, the microscopic origin of this universal enhancement of the ANE in the metallic multilayers still remains to be clarified. One of the unconventional behaviors is that the ANE coefficients in our Pt/Fe, Au/Fe, and Cu/Fe multilayers are opposite in sign to the coefficient for bulk Fe. In fact, we checked that the sign of the ANE signal in a 10-nm-thick plain Fe film on the GGG substrate is opposite to that in the multilayer samples (see black square data points in Fig. \ref{fig:3}), where the Fe film was prepared under the same condition as that for the multilayers. Since the parameters changing in our paramagnet/Fe multilayers are the Fe thickness and the interface density, the above experimental results suggest the presence of extraordinary thickness dependence of the ANE in the Fe films attached to the paramagnets or unconventional ANE at the paramagnet/Fe interfaces. In metallic multilayer systems, additional ANEs might be driven by the modulation of spin-orbit interaction due to the following candidates: (1) surface roughness of the Fe layers \cite{Zhou}, (2) interdiffusion and alloying at the paramagnet/Fe interfaces, (3) strain in the Fe layers, and (4) crystal-symmetry breaking at the paramagnet/Fe interfaces \cite{JACBland}. We believe that, at least for the ANE in the PM configuration, the candidate (1) is plausible to explain the universal enhancement of the ANE in the paramagnet/ferromagnet multilayers, because the others may depend on the species of the paramagnetic layers. Nevertheless, more systematic studies are required for the full understanding of the observed behavior. \par
In conclusion, we measured the anomalous Nernst effect (ANE) in the alternately-stacked Pt/Fe, Au/Fe, and Cu/Fe multilayer films in the in-plane and perpendicularly magnetized configurations to investigate the possible contribution of the ANE induced by magnetic proximity effects in the paramagnetic metals. The ANE in these multilayer systems was observed to be enhanced with increasing the layer density of the paramagnet/Fe interfaces in both the configurations irrespective of the presence or absence of the proximity ferromagnetism, although the origin of the ANE enhancement is yet to be revealed. Recently, the enhancement of the longitudinal spin Seebeck effect (LSSE) has also been demonstrated using Pt/Fe$_3$O$_4$ multilayer films \cite{SSE-multilayer}, metallic paramagnet/ferromagnet multilayers \cite{ANE-multilayer1}, and in magnetic oxide superlattices \cite{Shiomi-multilayer} in the in-plane magnetized configuration. These experimental results show the usefulness of magnetic multilayer systems for thermoelectric applications. We anticipate that further investigation of the ANE and LSSE in multilayer systems will lead to the establishment of novel mechanisms for enhancing thermoelectric and thermo-spin effects. \par
K.U., T.K., and T.S. contributed equally to this work. The authors thank Y. Shiomi, Y. Sakuraba, and J. Okabayashi for valuable discussions. This work was supported by PRESTO ``Phase Interfaces for Highly Efficient Energy Utilization'' from JST, Japan, Grant-in-Aid for Challenging Exploratory Research (26600067), Grant-in-Aid for Scientific Research on Innovative Areas ``Nano Spin Conversion Science'' (26103005), Grant-in-Aid for Scientific Research (A) (15H02012), Grant-in-Aid for Scientific Research (S) (25220910) from MEXT, Japan, and NEC Corporation. T.K. and T.O. are supported by JSPS through a research fellowship for young scientists. \par

\begin{thebibliography}{99}
%
%
\bibitem{ANE_Berger} L. Berger, Phys. Rev. B {\bf 5}, 1862 (1972).
\bibitem{ANE_Miyasato} T. Miyasato, N. Abe, T. Fujii, A. Asamitsu, S. Onoda, Y. Onose, N. Nagaosa, and Y. Tokura, Phys. Rev. Lett. {\bf 99}, 086602 (2007).
\bibitem{ANE_Mizuguchi} M. Mizuguchi, S. Ohata, K. Uchida, E. Saitoh, and K. Takanashi, Appl. Phys. Express. {\bf 5}, 093002 (2012).
\bibitem{ANE_Weischenberg} J. Weischenberg, F. Freimuth, S. Bl\"ugel, and Y. Mokrousov, Phys. Rev. B {\bf 87}, 060406(R) (2013).
\bibitem{ANE_Sakuraba} Y. Sakuraba, K. Hasegawa, M. Mizuguchi, T. Kubota, S. Mizukami, T. Miyazaki, and K. Takanashi, Appl. Phys. Express. {\bf 6}, 033003 (2013). 
\bibitem{spintronics2} S. R. Boona, R. C. Myers, and J. P. Heremans, Energy Environ. Sci. {\bf 7}, 885 (2014). 
\bibitem{ANE_Ramos} R. Ramos, M. H. Aguirre, A. Anad\ifmmode \acute{o}\else \'{o}\fi{}n, J. Blasco, I. Lucas, K. Uchida, P. A. Algarabel, L. Morell\ifmmode \acute{o}\else \'{o}\fi{}n, E. Saitoh, and M. R. Ibarra, Phys. Rev. B  {\bf 90}, 054422 (2014).  
%
%
\bibitem{spintronics1} S. Maekawa, H. Adachi, K. Uchida, J. Ieda, and E. Saitoh, J. Phys. Soc. Jpn. {\bf 82}, 102002 (2013).
%
%
\bibitem{TSSE_Uchida2008} K. Uchida, S. Takahashi, K. Harii, J. Ieda, W. Koshibae, K. Ando, S. Maekawa, and E. Saitoh, Nature {\bf 455}, 778 (2008). 
\bibitem{TSSE_Uchida2010} K. Uchida, J. Xiao, H. Adachi, J. Ohe, S. Takahashi, J. Ieda, T. Ota, Y. Kajiwara, H. Umezawa, H. Kawai, G. E. W. Bauer, S. Maekawa, and E. Saitoh, Nat. Mater. {\bf 9}, 894 (2010). 
\bibitem{TSSE_Jaworski2010} C. M. Jaworski, J. Yang, S. Mack, D. D. Awschalom, J. P. Heremans, and R. C. Myers, Nat. Mater. {\bf 9}, 898 (2010). 
%
%
\bibitem{LSSE_Uchida2010} K. Uchida, H. Adachi, T. Ota, H. Nakayama, S. Maekawa, and E. Saitoh, Appl. Phys. Lett. {\bf 97}, 172505 (2010).  
\bibitem{LSSE_Du2013PRL} D. Qu, S. Y. Huang, J. Hu,  R. Wu, and C. L. Chien, Phys. Rev. Lett. {\bf 110}, 067206 (2013). 
\bibitem{LSSE_Kikkawa2013PRL} T. Kikkawa, K. Uchida, Y. Shiomi, Z. Qiu, D. Hou, D. Tian, H. Nakayama, X.-F. Jin, and E. Saitoh, Phys. Rev. Lett. {\bf 110}, 067207 (2013). 
\bibitem{LSSE_Ramos2013} R. Ramos, T. Kikkawa, K. Uchida, H. Adachi, I. Lucas, M. H. Aguirre, P. Algarabel, L. Morell\ifmmode \acute{o}\else \'{o}\fi{}n, S. Maekawa, E. Saitoh, and M. R. Ibarra, Appl. Phys. Lett. {\bf 102}, 072413 (2013). 
\bibitem{LSSE_Uchida2013PRB} K. Uchida, T. Nonaka, T. Kikkawa, Y. Kajiwara, and E. Saitoh, Phys. Rev. B {\bf 87}, 104412 (2013). 
\bibitem{LSSE_Schreier2013} M. Schreier, A. Kamra, M. Weiler, J. Xiao, G. E. W. Bauer, R. Gross, and S. T. B. Goennenwein, Phys. Rev. B {\bf 88}, 094410 (2013).  
\bibitem{LSSE_Kikkawa2013PRB} T. Kikkawa, K. Uchida, S. Daimon, Y. Shiomi, H. Adachi, Z. Qiu, D. Hou, X.-F. Jin, S. Maekawa, and E. Saitoh, Phys. Rev. B {\bf 88}, 214403 (2013).  
\bibitem{LSSE_Rezende2014} S. M. Rezende, R. L. Rodr\ifmmode \acute{i}\else \'{i}\fi{}guez-Su\ifmmode \acute{a}\else \'{a}\fi{}rez, R. O. Cunha, A. R. Rodrigues, F. L. A. Machado, G. A. Fonseca Guerra, J. C. Lopez Ortiz, and A. Azevedo, Phys. Rev. B {\bf 89}, 014416 (2014). 
\bibitem{LSSE-UchidaJPCM} K. Uchida, M. Ishida, T. Kikkawa, A. Kirihara, T. Murakami, and E. Saitoh, J. Phys.: Condens. Matter {\bf 26}, 343202 (2014). 
%
%
\bibitem{Huang_2012PRL} S. Y. Huang, X. Fan, D. Qu, Y. P. Chen, W. G. Wang, J. Wu, T. Y. Chen, J. Q. Xiao, and C. L. Chien, Phys. Rev. Lett. {\bf 109}, 107204 (2012). 
\bibitem{Lu_2013PRL} Y. M. Lu, Y. Choi, C. M. Ortega, X. M. Cheng, J. W. Cai, S. Y. Huang, L. Sun, and C. L. Chien, Phys. Rev. Lett. {\bf 110}, 147207 (2013). 
%
\bibitem{Ibach} H. Ibach and H. L\"uth, {\it Solid-State Physics: An Introduction to Principles of Materials Science} (Springer, Berlin, 2009). 
\bibitem{DOS} D. A. Papaconstantopoulos, {\it Handbook of the Band Structure of Elemental Solids} (Plenum Press, New York, 1986).
%
\bibitem{ANE_Guo} G. Y. Guo, Q. Niu, and N. Nagaosa, Phys. Rev. B {\bf 89}, 214406 (2014).
%
%
\bibitem{Geprags_2012APL} S. Gepr\"ags, S. Meyer, S. Altmannshofer, M. Opel, F. Wilhelm, A. Rogalev, R. Gross, and S. T. B. Goennenwein, Appl. Phys. Lett. {\bf 101}, 262407 (2012). 
%
%
\bibitem{SSE-multilayer} R. Ramos, T. Kikkawa, M. H. Aguirre, I. Lucas, A. Anad\ifmmode \acute{o}\else \'{o}\fi{}n, T. Oyake, K. Uchida, H. Adachi, J. Shiomi, P. A. Algarabel, L. Morell\ifmmode \acute{o}\else \'{o}\fi{}n, S. Maekawa, E. Saitoh, and M. R. Ibarra, arXiv:1503.05594 (2015).
\bibitem{ANE-multilayer1} K.-D. Lee, D.-J. Kim, H. Y. Lee, S.-H. Kim, J.-H. Lee, K.-M. Lee, J.-R. Jeong, K.-S. Lee, H.-S. Song, J.-W. Sohn, S.-C. Shin, and B.-G. Park, arXiv:1504.00642 (2015). 
%
%
\bibitem{comment-PSSE} Recently, the SSE was found to appear even in paramagnetic metal/paramagnetic insulator junction systems \cite{paramagnetic-SSE}. This paramagnetic SSE may modulate the normal ($H$-linear) component of the transverse thermoelectric voltage at low temperatures but does not affect the anomalous component of the voltage.  
\bibitem{paramagnetic-SSE} S. M. Wu, J. E. Pearson, and A. Bhattacharya, Phys. Rev. Lett. {\bf 114}, 186602 (2015). 
%
%
\bibitem{ISHE1} A. Azevedo, L. H. Vilela-Le\~ao, R. L. Rodr\ifmmode \acute{i}\else \'{i}\fi{}guez-Su\ifmmode \acute{a}\else \'{a}\fi{}rez, A. B. Oliveira, and S. M. Rezende, J. Appl. Phys. {\bf 97}, 10C715 (2005). 
\bibitem{ISHE2} E. Saitoh, M. Ueda, H. Miyajima, and G. Tatara, Appl. Phys. Lett. {\bf 88}, 182509 (2006). 
\bibitem{ISHE3} S. O. Valenzuela and M. Tinkham, Nature {\bf 442}, 176 (2006). 
%
%
\bibitem{comment-PM} In the PM configuration, a possible small temperature gradient perpendicular to the film plane does not affect the voltage signal, since the Nernst voltage is not generated due to the collinear orientation of the perpendicular temperature gradient and the magnetization or magnetic field. 
\bibitem{comment-ANEcoefficient} $S_{\rm ANE}$ shown in Eq. (\ref{equ:ANE1}) and $S_{\rm 0}$ are the transverse electric fields normalized by the temperature gradients in the metallic film and in the GGG substrate, respectively. Therefore, $S_{\rm 0}$ is different from $S_{\rm ANE}$ in the IM configuration because of the difference in the thermal conductivity between the film and the substrate, while they are equivalent to each other in the PM configuration \cite{LSSE_Kikkawa2013PRB,LSSE-UchidaJPCM}. 
%
%
\bibitem{XMCD_FeCu} S. Pizzini, A. Fontaine, C. Giorgetti, E. Dartyge, J.-F. Bobo, M. Piecuch, and F. Baudelet, Phys. Rev. Lett. {\bf 74}, 1470 (1995). 
\bibitem{XMCD_FePt} W. J. Antel, Jr., M. M. Schwickert, T. Lin, W. L. O'Brien, and G. R. Harp, Phys. Rev. B {\bf 60}, 12933 (1999).
\bibitem{Calc_FePt_FeAu} R. Tyer, G. van der Laan, W. M. Temmerman, Z. Szotek, and H. Ebert, Phys. Rev. B {\bf 67}, 104409 (2003). 
\bibitem{Calc_FeCu} K. Hirai, Physica B {\bf 345}, 209 (2004).  
\bibitem{XMCD_CoAu} F. Wilhelm, M. Angelakeris, N. Jaouen, P. Poulopoulos, E. Th. Papaioannou, Ch. Mueller, P. Fumagalli, A. Rogalev, and N. K. Flevaris, Phys. Rev. B {\bf 69}, 220404(R) (2004).  
%
%
\bibitem{Zhou} L. Zhou, V. L. Grigoryan, S. Maekawa, X. Wang, and J. Xiao, Phys. Rev. B {\bf 91}, 045407 (2015). 
\bibitem{JACBland} J. A. C. Bland and B. Heinrich (Eds.), {\it Ultrathin Magnetic Structures I: An Introduction to the Electronic, Magnetic and Structural Properties} (Springer, Berlin, 2005). 
%
%
\bibitem{Shiomi-multilayer} Y. Shiomi, Y. Handa, T. Kikkawa, and E. Saitoh, arXiv:1505.01991 (2015). 
%
%
\end{thebibliography}
\end{document}